\documentclass[twocolumn]{aastex631}

\usepackage{graphicx}	
\usepackage{amsmath}	
\usepackage{bm}
\usepackage{amssymb}	
\usepackage{color}
\usepackage{ulem}
\usepackage{soul}

\newcommand \mb[1] {{\mathbf{#1}}}

\begin{document}

\title{Stochastic Wave Dark Matter with Fermi-LAT $\gamma$-ray Pulsar Timing Array}

\author[0000-0001-9483-1099]{Hoang Nhan Luu}
\affiliation{Department of Physics and Jockey Club Institute for Advanced Study, \\
The Hong Kong University of Science and Technology, Hong Kong S.A.R., China}
\affiliation{DIPC, Basque Country UPV/EHU, San Sebastian, E-48080, Spain}
\email{hnluu@connect.ust.hk}

\author[0000-0002-5248-5076]{Tao Liu}
\affiliation{Department of Physics and Jockey Club Institute for Advanced Study, \\
The Hong Kong University of Science and Technology, Hong Kong S.A.R., China}
\email{taoliu@ust.hk}

\author[0000-0002-2116-4659]{Jing Ren}
\affiliation{Institute of High Energy Physics, Chinese Academy of Sciences, Beijing 100049, China}
\email{renjing@ihep.ac.cn}

\author[0000-0002-8785-8979]{Tom Broadhurst}
\affiliation{University of the Basque Country UPV/EHU, Department of Theoretical Physics, Bilbao, E-48080, Spain}
\affiliation{DIPC, Basque Country UPV/EHU, San Sebastian, E-48080, Spain}
\affiliation{Ikerbasque, Basque Foundation for Science, Bilbao, E-48011, Spain}
 
\author[0000-0001-5801-2547]{Ruizhi Yang}
\affiliation{Deep Space Exploration Laboratory/School of Physical Sciences, \\
CAS Key Laboratory for Research in Galaxies and Cosmology/Department of Astronomy \\ School of Astronomy and Space Science, University of Science and Technology of China, Hefei 230026, China}

% \author[0000-0002-2662-6912]{Jie-Shuang Wang}
\author[0000-0002-2662-6912]{Jie-Shuang Wang}
\affiliation{Max-Planck-Institut f\"ur Kernphysik, Saupfercheckweg 1, D-69117 Heidelberg, Germany}

\author[0009-0005-6541-214X]{Zhen Xie}
\affiliation{Deep Space Exploration Laboratory/School of Physical Sciences, \\
CAS Key Laboratory for Research in Galaxies and Cosmology/Department of Astronomy  \\ School of Astronomy and Space Science, University of Science and Technology of China, Hefei 230026, China}
 
\begin{abstract}
Pulsar timing arrays (PTAs) can detect disturbances in the fabric of spacetime on a galactic scale by monitoring the arrival time of pulses from millisecond pulsars (MSPs). Recent advancements have enabled the use of $\gamma$-ray radiation emitted by MSPs, in addition to radio waves, for PTA experiments. Wave dark matter (DM), a prominent class of DM candidates, can be detected with PTAs due to its periodic perturbations of the spacetime metric. In response to this development, we perform in this Letter a first analysis of applying the $\gamma$-ray PTA to detect the ultralight axion-like wave DM, with the data of Fermi Large Area Telescope (Fermi-LAT). Despite its much smaller collecting area, the Fermi-LAT $\gamma$-ray PTA demonstrates a promising sensitivity potential. We show that the upper limits not far from those of the dedicated radio-PTA projects can be achieved. Moreover, we initiate a cross-correlation analysis using the data of two Fermi-LAT pulsars. The cross-correlation of phases, while carrying key information on the source of the spacetime perturbations, has been ignored in the existing data analyses for the wave DM detection with PTAs. Our analysis indicates that taking this information into account can improve the sensitivity to wave DM by $\gtrsim 50\%$ at masses below $10^{-23}\,$eV.
\end{abstract}

\keywords{Wave dark matter --- Pulsar timing array --- Fermi-LAT}

\section{Introduction}

Wave dark matter (DM) represent a class of bosonic DM candidates with a mass $ \lesssim 1\,$eV (for a review on wave DM, see, e.g.,~\citealt{Hui:2021tkt}). Such a candidate has a large occupation number at a scale of de Broglie wavelength in the environment of Milky Way galaxy and hence demonstrates a coherent wave nature. Wave DM with a mass $\lesssim 10^{-18}$~eV is usually dubbed ``ultralight''~\citep{Marsh:2015xka}, for its astronomical-scale de Broglie wavelength. The ``Fuzzy" DM~\cite[FDM,][]{Hu:2000ke,Schive:2014dra,Hui:2016ltb}, as one special class of ultralight wave DM with a mass $\sim 10^{-22}$~eV, is of particular interest in astronomy. As its de Broglie wavelength is comparable to the size ($\sim 0.5$~kpc) of dwarf-galaxy dark cores, the FDM can be applied to address several small-scale problems on astronomical structures, such as core-cusp problem~\citep{deBlok:2010} (namely a contradiction between the predicted Navarro-Frenk-White ``cusps" for the DM halo profile~\citep{Navarro:1995iw} in N-body simulations of standard collisionless cold DM, and the observation of  approximately constant DM density in the central parts of galaxies) which are challenging the standard cold DM scenario.

Ultralight wave DM, including FDM, could be detected by employing its impacts on either the high-$z$ Universe~\citep{Marsh:2015xka}, such as the formation of large-scale structure, or the astronomy of local galaxy, e.g., the timing and polarization of pulsar pulses. The successful launching of James Webb Space Telescope provides a great tool for achieving the goal with the first method, while the active and upcoming pulsar timing programs enable us to explore this puzzle with the second one.

Given that pulsars can emit electromagnetic pulses with extraordinary regularity, a Pulsar Timing Array~\citep[PTA,][]{Sazhin:1978, Detweiler:1979, Hellings:1983fr} consisting of broadly distributed and well-timed millisecond pulsars has been proposed as an astronomical interferometer to detect the gravitational waves (GWs) of $10^{-9}-10^{-6}$~Hz. In 2013, Khmelnitsky and Rubakov pointed out that the wave DM can periodically perturb galactic metric, at a level of $\mathcal O(v^2)$, and hence be detected with the PTA~\citep{Khmelnitsky:2013lxt}. The concept of Pulsar Polarization Array (PPA) is much newer~\citep{Liu:2021zlt}. This new methodology was developed recently to reveal new phenomena in astronomy with a common correlated polarization signal, such as the one caused by ultralight axion-like wave DM~\citep{Liu:2021zlt}. Because its Chern-Simons coupling with electromagnetic field strength is parity-violating, the axion-like wave DM can modulate the position angle of linearly polarized pulsar light while the light travels across the DM halo. The PPA is thus highly suited for its detection~\citep{Liu:2021zlt}. Notably, the PTA and PPA methods here are based on gravitational and non-gravitational effects, respectively. So combining them can further strengthen the capability of arrayed pulsars for exploring the nature of DM.  

Due to the good sensitivity of large radio telescopes, the PTAs have been mostly realized by monitoring radio millisecond pulsars. In addition to the ongoing programs such as the Parkes PTA~\cite[PPTA,][]{Manchester:2013}, the European PTA~\cite[EPTA,][]{Desvignes:2016}, the Chinese PTA~\cite[CPTA,][]{Lee:2016} and the North American Nanohertz Observatory for Gravitational Waves~\cite[NANOGrav,][]{NANOGrav:2015} PTA, in the near future the international Square Kilometer Array radio telescope~\citep{Dewdney:2009} is also expected to underpin the construction of global PTA. These dedicated efforts were extended to a higher-frequency band recently. With the data accumulated by Fermi Large Area Telescope (Fermi-LAT), the first $\gamma$-ray PTA ($\gamma$-PTA) constraints for the stochastic GWs have been delivered~\citep{Fermi-LAT:2022wah}. While its sensitivity is limited by statistics, such a high-frequency PTA benefits from a suppression of intrinsic red noise of dispersion measure variance and relative weak effects from white noises such as radio interference and jitter. The $\gamma$-PTA thus well-complements the radio one. Such a ``radio + high-frequency'' strategic development could be  extended to the PPA measurements in the near future, with the launch of dedicated X-ray polarimetry programs such as the missions of Imaging X-ray Polarimetry Explorer~\cite[IXPE,][]{IXPE1} and Enhanced X-ray Timing and Polarimetry~\cite[eXTP,][]{eXTP:2018anb}. For these telescopes, a precision $\sim \mathcal O(1^\circ)$ can be achieved for measuring polarization angle of bright sources such as bright pulsars~\citep{Doroshenko:2023} and accretion flows around black holes~\citep{DeRosa:2019}.

Given its high scientific value, in this Letter we will demonstrate a first application of high-frequency PTA for detecting the ultralight axion-like wave DM, with the Fermi-LAT data. 

%\section{PBP-Based Methodology for Detecting Stochastic Wave DM}

\section{Methodology for Detecting Stochastic Wave DM}

The profile of stochastic wave DM can be described as an uncorrelated superposition of particle waves. For a proof of concept, let us consider ultralight axion-like wave DM. Its profile can be put as
\begin{eqnarray}\label{eq:axionf}
a(\mathbf{x},t) \approx \frac{\sqrt{\rho(\mathbf{x})}}{m_a}\sum_{\mathbf{v}\in\Omega}C_{\mb{v}}\cos[m_a(t-\mathbf{v}\cdot\mathbf{x})+\psi_{\mathbf{v}}]. 
\end{eqnarray}
Here $C_{\mb{v}}$ is equal to $\alpha_{\mb{v}}\sqrt{f(\mb{v})}(\Delta v)^{3/2}$ with $\mb{v}\in\Omega$ denoting lattice sites in phase space and $\Delta v$ being their spacing. $\rho(\mb{x})$ and $f(\mathbf{v})$ represent energy density profile and velocity distribution. 
The stochastic nature of wave DM is then encoded as a random amplitude
$\alpha_\mathbf{v}\in(0,+\infty)$ and phase $\psi_\mathbf{v}$, which follow Rayleigh distribution $p(\alpha_\mathbf{v})=\alpha_\mathbf{v} e^{-\alpha_\mathbf{v}^2/2}$ and uniform distribution respectively (for a proof, see Sec.~II in~\citealt{Foster:2017hbq}).

Because of its gradient energy, the wave-DM halo in a galaxy can induce an oscillating gravitational potential $\Psi(\mb{x},t)$ in proportion to its local density. This further yields a shift to the time of arrival (TOA) of pulsar pulses~\citep{Khmelnitsky:2013lxt}:
\begin{eqnarray}\label{eq:dti1}
\Delta t_p(t) 
&=& \int_{t-L_p}^t \left[\Psi(\mb{x}_e,t)-\Psi(\mb{x}_p,t')\right]dt'\nonumber\\
&=&\frac{\pi G}{4m_a^3}
\sum_{\mb{v},\mb{v'}} C_{\mb{v}}C_{\mb{v'}}
\Big(\rho_e\sin[2m_a t+\varphi^{(e)}_{\mb{v}\mb{v}'}] 
 \nonumber\\
&&-\rho_p\sin[2m_a t+\varphi^{(p)}_{\mb{v}\mb{v}'}] \Big)\,,
\label{Eq:phase_axion}
\end{eqnarray}
where $L_p$ is a distance from the $p$-th pulsar to the Earth, $\mb{x}_{i}$ denotes the position of the Earth ($i=e$) and pulsar ($i=p$), and $\rho_i=\rho(\mb{x}_i)$ is the DM energy density at $\mb{x}_i$. $\Delta t_p(t) $ is then determined as a difference between an “Earth” and a “pulsar” terms. These terms oscillate with a frequency  $\sim 2 m_a$, with the frequency being set by the mass of axion-like particles $m_a$. For the oscillation phases, the spatial dependence on pulsars is encoded in $\varphi^{(e)}_{\mb{v}\mb{v}'}=\psi_{\mb{v}}+\psi_{\mb{v'}}$ and 
$\varphi^{(p)}_{\mb{v}\mb{v}'}=-2m_a L_p-m_a(\mb{v}+\mb{v'})\cdot(\mb{x}_p-\mb{x}_e)+\psi_{\mb{v}}+\psi_{\mb{v'}}$. Note, both $\varphi^{(e)}_{\mb{v}\mb{v}'}$ and $\varphi^{(p)}_{\mb{v}\mb{v}'}$ contain two random phases due to a quadratic dependence of $\Delta t_p(t)$ on the wave DM profile $a(\mb{x},t)$. This is different from the polarization signal of axion-like wave DM where such a dependence is linear~\citep{Liu:2021zlt}.

The radio-PTA analysis is usually performed with a TOA-based method, which involves the comparison
of a time-referenced model of the pulse profile with the observed data. For many of its pulsars, the Fermi-LAT can only reconstruct a few TOAs per year due to the low statistics of received $\gamma$-ray photons from these pulsars. Thus, its ability to detect short-time-scale ($f\gtrsim 1~{\rm yr}^{-1}$) signals gets highly constrained in this case. A photon-by-photon (PBP)-based method, which we will take below, thus has been introduced in~\cite{Fermi-LAT:2022wah} to address this limitation.  

At the Fermi-LAT~\citep{Atwood:2009}, the $\gamma$-ray data are recorded in terms of energy $E_i$ and arrival time $t_i$ for the $i$-th photon. The relevant physical observable is pulsar spin phase (PSP) at certain photon arrival time, {\it i.e.,} $0\leq \phi(t_i)<1$. 
The evolution of PSP can be predicted with a timing model. For the $p$-th pulsar, it is defined as
	\begin{align}
		\phi^{(p)}(t) = \phi^{(p)}_{\rm tm}(t) + \delta\phi^{(p)}_{\rm tn} (t) + \delta\phi^{(p)}_a(t)  \, .
	\end{align}
    There are in total three classes of contributions to $\phi^{(p)}$. $\phi^{(p)}_{\rm tm}$ denotes the contributions from a deterministic timing model, encoding the effects of pulsar spindown, position, proper motion, etc. For example, the spindown effect yields a contribution $\phi_0^{(p)} + \nu^{(p)} t + \frac{1}{2}\dot{\nu}^{(p)}t^2 + \mathcal{O}(t^3)$, with $\nu$ and $\dot{\nu}$ being pulsar spin frequency and its spin down rate, respectively.

    $\delta\phi_{\rm tn}^{(p)}$ denotes the contributions arising from potential timing-noise processes such as red spin noise. It can be written as in terms of Fourier modes~\citep{Fermi-LAT:2022wah}: 		
    \begin{equation}
    \begin{split}
&     \delta\phi^{(p)}_{\rm tn} (t) = \sqrt{\frac{2}{T}} \nu^{(p)}_0 \times  \\
    & \left[ \sum_k \beta^{(p)}_{2k} \cos\left( 2\pi k \frac{t}{T} \right) + \beta^{(p)}_{2k+1} \sin\left( 2\pi k \frac{t}{T}  \right)  \right ]\, . \label{ansatz}
    \end{split}
      \end{equation}
Here, $T\approx 12.5\,$yrs represents the total observation time, which is approximately equal for all pulsars. $\nu_0^{(p)}$ denotes the intrinsic spin frequency, and $k$'s are the indices of Fourier amplitudes. The red spin noise can be  modeled as a Gaussian  process, with a covariant matrix element for the  Fourier mode $\beta_k$
 \begin{eqnarray}
 \langle\beta^{(p)}_k \beta^{(q)}_l\rangle =\frac{1}{2}\delta_{pq}\delta_{kl}P_{\rm tn}(f_k)\,,
\label{Eq:power_spectrum}
 \end{eqnarray}
 where $P_{\rm tn}(f) = \frac{A_{\rm tn}^2}{12\pi^2}\left( f/{\rm yr}^{-1} \right)^{-\Gamma_{\rm tn}}{\rm yr}^3$ denotes the power spectrum, with $A_{\rm tn}$ and $\Gamma_{\rm tn}$ being its amplitude and spectral index, respectively, and $f_k=k/T$ is the frequency of $k$-th Fourier mode. At last, $\delta\phi^{(p)}_a(t)$ denotes the PSP deviation induced by a signal such as the stochastic wave DM. It has the same format as that of $\delta\phi^{(p)}_{\rm tn} (t)$ in Eq.~(\ref{ansatz}).   
    
The $\delta\phi^{(p)}_a(t)$ is related to the signal $\Delta t_p(t)$ in  Eq.~(\ref{eq:dti1}) as $\delta\phi^{(p)}_a(t)=\nu_0^{(p)}\Delta t_p(t)$, with $\Delta t_p=A_{c}^{(p)}\cos(2m_a t) + A_{s}^{(p)}\sin(2m_a t)$.  
This results in 
 \begin{eqnarray}    
 \delta\phi^{(p)}_a(t)=\nu_0^{(p)}A_{c}^{(p)}\cos(2m_a t) + \nu_0^{(p)}A_{s}^{(p)}\sin(2m_a t) \, ,
    \end{eqnarray}
where
\begin{equation}
\begin{split}
A_{c}^{(p)}&= \frac{\pi G}{4m_a^3}
\sum_{\mb{v},\mb{v'}} 
C_{\mb{v}}C_{\mb{v'}}
\Big[\rho_p\sin\varphi^{(p)}_{\mb{v}\mb{v}'}-
\rho_e\sin\varphi^{(e)}_{\mb{v}\mb{v}'}\Big] 
 \, ,  \\ 
A_{s}^{(p)}&= \frac{\pi G}{4m_a^3}
\sum_{\mb{v},\mb{v'}} C_{\mb{v}}C_{\mb{v'}}
\Big[\rho_p\cos\varphi^{(p)}_{\mb{v}\mb{v}'}-
\rho_e\cos\varphi^{(e)}_{\mb{v}\mb{v}'}\Big] \, . 
\label{Eq:Ac_As_amplitude}
\end{split}
\end{equation}
These coefficients follow a multivariate Gaussian distribution with zero mean. Assuming that the DM velocity is isotropic and peaks sharply at the virial velocity of Milky Way, {\it i.e.}, $f(\mb{v})\approx (4\pi v^2)^{-1}\delta(v-v_0)$ with $v_0\sim 10^{-3}$, the covariance matrix elements are given by
\begin{eqnarray}
\langle A_{c}^{(p)}A_{c}^{(q)}\rangle&=&
\langle A_{s}^{(p)}A_{s}^{(q)}\rangle\nonumber\\
&\approx& \left( \frac{\pi G}{2m_a^3}\right)^2
\Big[\rho_e^2+\rho_p\rho_q\cos[2m_a  L_{pq}]\frac{\sin^2 y_{pq}}{y^2_{pq}}\nonumber\\
&&-\rho_e\rho_p\cos[2m_a L_p]\frac{\sin^2 y_{ep}}{y^2_{ep}}-(p\to q)\Big]\,,\nonumber\\
\langle A_{s}^{(p)}A_{c}^{(q)}\rangle
&\approx&\left( \frac{\pi G}{2m_a^3}\right)^2
\Big[\rho_p\rho_q\sin[2m_a L_{pq}]\frac{\sin^2 y_{pq}}{y^2_{pq}}\nonumber \\
&&+\rho_e\rho_p\sin[2m_a L_p]\frac{\sin^2 y_{ep}}{y^2_{ep}}-(p\to q)\Big]\,  ,\label{Eq:cov_term}
\end{eqnarray}
with $L_{pq}=L_{p}-L_q$ and $y_{ij}=|\mb{x}_i-\mb{x}_j|/l_c$. Here $l_c\equiv 1/(m_a v_0)$ is the coherence length of wave DM. $\sin y_{ij}/y_{ij}$ is a sinc function, encoding the spatial correlation of signal. To derive Eq.~(\ref{Eq:Ac_As_amplitude}), 
$\langle\alpha^2_{\mb{v}}\rangle=2$ and $\langle\sin(\alpha+\psi_{\mb{v_1}}+\psi_{\mb{v_1'}})\sin(\beta+\psi_{\mb{v_2}}+\psi_{\mb{v_2'}})\rangle=\frac{1}{2}\cos(\alpha-\beta)\left[\delta_{\mb{v_1},\mb{v_2}}\delta_{\mb{v'_1},\mb{v'_2}}+
\delta_{\mb{v_1},\mb{v'_2}}\delta_{\mb{v'_1},\mb{v_2}}\right]$ have been applied. Compared to the stochastic GWs with a coherence length $l_c \sim 1/\omega$, the wave DM of the same frequency has a much longer $l_c$ due to its non-relativistic character. The ``pulsar'' term thus plays a more important role in cross-correlating the signals~\citep{Liu:2021zlt}. This significance can be further enhanced for the pulsars close to the galactic center where the DM halo is dense~\citep{DeMartino:2017qsa, Liu:2021zlt}.
    
For a PTA with $\mathcal{N}$ pulsars, the log-likelihood function in the PBP method is given by~\cite{Fermi-LAT:2022wah}
\begin{eqnarray}
 \ln && \mathcal{L} = \sum_{p, i} \ln \left( w^{(p)}_i f^{(p)}\left[\phi^{(p)}_i\right] + 1 - w^{(p)}_i \right) \\ &&- 0.5\left( \beta_{\rm tn}^T \mathcal{C}_{\rm tn}^{-1}\beta_{\rm tn} + \ln|\mathcal{C}_{\rm tn}| + \beta_a^T \mathcal{C}_a^{-1}\beta_a +\ln|\mathcal{C}_a| \right)\, . \nonumber  \label{Eq:single_loglik}
\end{eqnarray}  
Here the terms of first line describe a weighted Poisson distribution of PSP with a summation over the photons from all pulsars, $f^{(p)}$ denotes a template known in prior fitting the data, and $w^{(p)}_i$ measures the probability for a $i$-th photon to originate from the $p$-th pulsar. The rest is simply the Gaussian log-likelihood for random noise and signal processes. The terms of $\mathcal{C}_{\rm tn}$ arise from red spin noise, with 
\begin{align}
&(\mathcal{C}_{\rm tn})_{p,k;q,l} = \delta_{pq}\delta_{kl}P_{\rm tn}(f_k)/2 \, ,\\
&\beta_{\rm tn}=(\beta^{(1)}_1,...,\beta^{(1)}_{N_1},..., \beta^{(\mathcal{N})}_1,...,\beta^{(\mathcal{N})}_{N_{\mathcal{N}}})^T  \, .
\end{align}
The terms of $\mathcal{C}_{\rm a}$ account for the contributions of  stochastic wave DM, with 
\begin{align} 
&(\mathcal{C}_a)_{pq}=\nu_0^{(p)}\nu_0^{(q)}
    \begin{pmatrix}
 	  \langle A^{(p)}_c A^{(q)}_c \rangle & \langle A^{(p)}_c A^{(q)}_s \rangle  \\
 	\langle A^{(p)}_s A^{(q)}_c \rangle &  \langle A^{(p)}_s A^{(q)}_s \rangle 
 	\end{pmatrix}\,, \label{eq:Cspq} \\
   &\beta_{a}=(\nu_0^{(1)}A^{(1)}_c,\nu_0^{(1)}A^{(1)}_s,...,\nu_0^{(\mathcal{N})}A^{(\mathcal{N})}_c,\nu_0^{(\mathcal{N})}A^{(\mathcal{N})}_s)^T    \, . \nonumber
\end{align}
Here the elements of $2\mathcal{N} \times 2\mathcal{N}$ covariant matrix $(\mathcal{C}_a)_{pq}$ are calculated in Eq.~(\ref{Eq:cov_term}). 
Cross-correlating the data of pulsars thus can help to distinguish the signals from the uncorrelated noises and identify the nature of signal, as the Hellings-Downs curve does for the GW detection~\citep{Hellings:1983fr}\footnote{For a toy-case demonstration on this point in detecting the axion-like wave DM with the PPA, please see Fig.~1 in~\cite{Liu:2021zlt}}.

\section{Sensitivity Analysis}

We analyze the Fermi-LAT $\gamma$-ray data for 28 out of 35 pulsars~\citep{Kerr:2022}, excluding J0312-0921, J0613-0200, J1513-2550, J1543-5149, J1741+1351, J1858-2216, J2034+3632\footnote{As these pulsars are relatively weak, including them into the analysis does not improve the upper limits remarkably, but significantly slows down the analysis.}, with the log-likelihood in Eq.~\eqref{Eq:single_loglik}. Since the distance of these pulsars to the Earth is small compared to the scale of DM density variations, we assume $\rho_p \simeq \rho_e$ in Eq.~\eqref{Eq:cov_term} and then absorb $\rho_e$ into a dimensionless parameter $\Psi_c=\pi G\rho_e/m_a^2$. Here $\Psi_c$ is an amplitude measure for the $\Psi(\mb{x},t)$ oscillation induced by the axion-like wave DM. It can be viewed as a counterpart of $h_c/f^{1/2}$, namely the GW characteristic strain with a frequency factor. To derive the upper limit on $\Psi_c$, we first analytically marginalize over the nuisance parameters $\phi_0$, $\nu$, $\dot{\nu}$, $\beta_{\rm tn}$ and $\beta_{a}$ to derive the marginal likelihood  $\mathcal{L}_m$, following the procedure in \cite{Fermi-LAT:2022wah}. Then, we numerically marginalize over the amplitude $A_{\rm tn}$ and  spectral index $\Gamma_{\rm tn}$ for the red spin noise, and the pulsar distance $\{L_p\}$ for $p=1,...,\mathcal{N}$ in $\mathcal{L}_m$\footnote{Since the line-of-sight direction of pulsars is well measured, the distance between  pulsars can be determined by their distances to the  Earth.}. The exclusion limits of $\Psi_c$ or $\rho_e$ at 95\% C.L. are finally computed with 
\begin{align}
		\int_0^{\Psi_{95}} d\Psi_c\, P(\Psi_c) \int d{\bf x} \,\mathcal{L}_m(\Psi_c, {\bf x}|m_a) P({\bf x}) = 0.95 , \label{Eq:Psi_limit}
\end{align}
where ${\bf x}=\{ A_{\rm tn}, \Gamma_{\rm tn}, L_p \}$ and $P({\bf x})$, $P(\Psi_c)$ denote the relevant priors. For the dimensionless parameter $\Psi_c$, we set a uniform prior with $\Psi_c\in {\rm U}\left[ 10^{-20}, 10^{-10} \right]$. For the red spin noise, we set $A_{\rm tn}\in {\rm U}\left[ 10^{-20}, 10^{-10} \right]$ and $\Gamma_{\rm tn}\in {\rm U}\left[ 1, 6 \right]$. These prior ranges have been chosen such that the associated PSP for the signals does not exceed the timing model contribution and hence can be treated perturbatively with respect to the timing model. We employ the package \texttt{PTMCMCSampler}~\citep{justin_ellis_2017_1037579} to perform the numerical marginalization on nuisance parameters for the cases where the timing noises and/or the  distance uncertainties are involved.

As a ``fiducial'' demonstration, let us consider the sensitivities with the terms independent of pulsar distance only in Eq.~\eqref{Eq:cov_term}, for both auto- and cross-correlations. The covariance matrix $(\mathcal{C}_a)_{pq}$ in Eq.~\eqref{eq:Cspq} is reduced to 
   \begin{align}
       (\mathcal{C}_a)_{pq}  \simeq \dfrac{\Psi_c^2 \, \nu_0^{(p)}\nu_0^{(q)}}{4 m_a^2}(1+\delta_{pq}) \begin{pmatrix}
 			1 & 0 \\
 			0 & 1 
 		\end{pmatrix}\,. \label{Eq:Cspp}
    \end{align}
 In this case, $\mathcal{L}_m$ is also independent of the pulsar distance, and the numerical marginalization of $\{L_p\}$ in Eq.~(\ref{Eq:Psi_limit}) becomes trivial. This fiducial signal model will be referred to as \textit{simplified correlation} below.

    \begin{figure}[htb]
		\centering
        \includegraphics[scale=0.56]{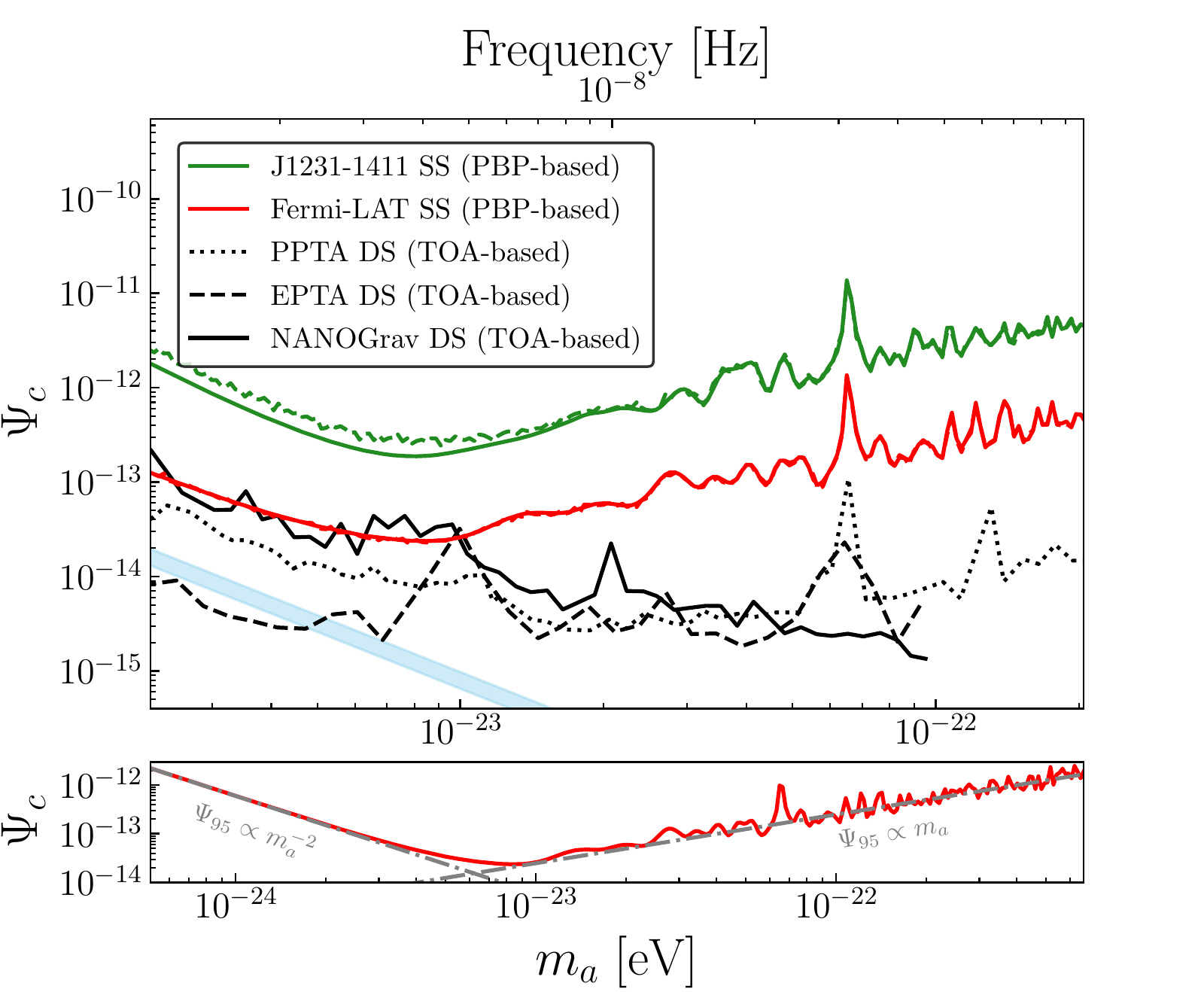} 
        \caption{Upper limits of the Fermi-LAT $\gamma$-PTA (95\% C.L.) on the dimensionless measure $\Psi_c$, as a function of $m_a$. ``SS'' and ``DS'' in the legend are abbreviations of ``stochastic signal'' and ``deterministic signal''. The limits are computed in the signal model with {\it simplified correlation.} The red dashed and solid curves denote the limits obtained with the Fermi-LAT data of 28 pulsars, with and without the timing noises, respectively. The best single-pulsar limits, which arise from pulsar J1231-1411, are shown as the green curves. As a reference, we present the most recent radio-PTA limits obtained with 
        the PPTA data \cite[black-dotted,][]{Porayko:2018sfa}, the EPTA data \cite[black-dashed,][]{EuropeanPulsarTimingArray:2023qbc}, and the NANOGrav data \cite[black-solid,][]{NANOGrav:2023hvm}. Additionally, we show the theoretical prediction for $\Psi_c$ as a light-blue band by setting $\rho_e$ to a value of local DM density~\citep{deSalas:2020hbh}. The lower sub-panel shows the same upper limits but with an extended mass range to highlight the scaling of $\Psi_{95}$ with $m_a$.}
	\label{Fig:exclusion1_DM}
    \end{figure}

     \begin{figure}[htb]       
        \includegraphics[scale=0.56]{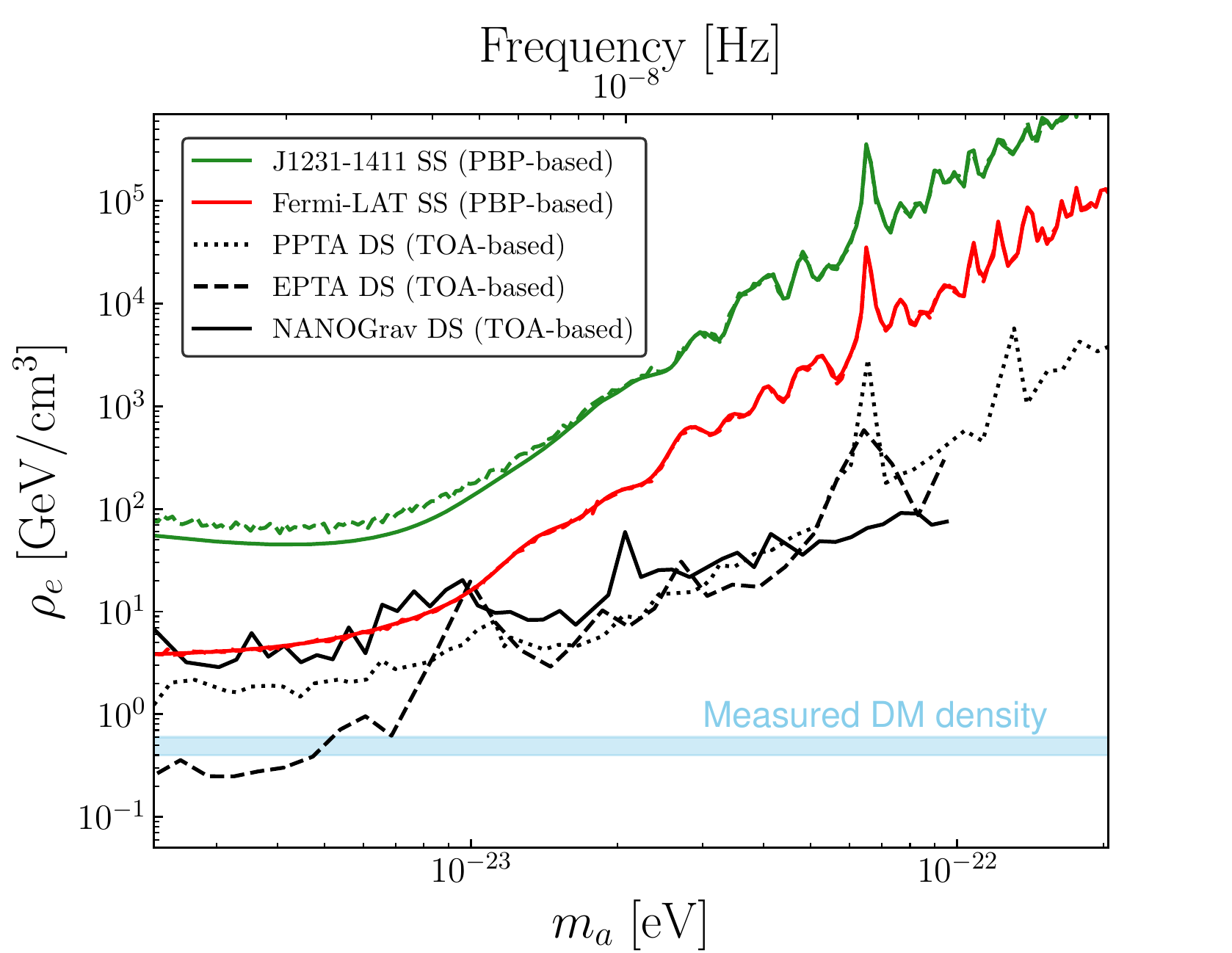}
		\caption{Upper limits of the Fermi-LAT $\gamma$-PTA (95\% C.L.) on the local DM density $\rho_e$, as a function of $m_a$. The description of curves is the same as that in Fig.~\ref{Fig:exclusion1_DM}. 
  }
		\label{Fig:exclusion2_DM}
    \end{figure}

With this fiducial signal model, we display the upper limits of the Fermi-LAT $\gamma$-PTA  on $\Psi_c$ and $\rho_e$ (95\% C.L.) in Fig.~\ref{Fig:exclusion1_DM} and~\ref{Fig:exclusion2_DM} respectively, as a function of the wave DM mass $m_a$. As the Fermi-LAT is not a dedicated PTA mission, its statistics for the PTA is relatively limited. The sensitivity potential demonstrated for the $\gamma$-PTA however is quite promising\footnote{We note that the limits obtained with stochastic signals actually are weaker compared to their deterministic counterparts~\citep{Centers:2019dyn} for an identical data set.}. For $m_a \lesssim 1/T \sim 10^{-23}~{\rm eV}$, these limits are comparable to the PPTA and NANOGrav ones~\citep{NANOGrav:2023hvm}. For larger $m_a$ values, they become approximately one order of magnitude weaker than the PPTA limits~\citep{Porayko:2018sfa}. Interestingly, as the TOA shift $\Delta t_p$ is $\propto \Psi_c$ for the axion-like wave DM and $\propto h_c/f^{1/2}$ for the GWs, these $\gamma$-PTA limits exhibit a scaling behavior with $m_a$ (or frequency) similar to those obtained with a TOA-based method on the characteristic strain~\citep{Hazboun:2019vhv} after factoring out $f^{1/2}$.  Approximately, we have $\Psi_{95} \propto m_a$ for $m_a > 1/T\sim 10^{-23}~{\rm eV}$, and $\Psi_{95} \propto m_a^{-2}$ for $m_a < 1/T$.

For both PTA and PPA detections, cross-correlating the data from the arrayed pulsars plays a crucial role in recognizing the nature of signals, which might be manifested as a specific pattern for the signal spatial correlation, or a distinguishable way for the correlated signals to depend on the relative position of these pulsars. For the axion-like wave DM, such an effect arises from a full consideration of its wave phase, or more explicitly the $\sim m_a \mathbf{v}\cdot\mathbf{x}$ term in this phase (see Eq.~(\ref{eq:axionf})), in calculating the covariance matrix $\mathcal{C}_a$ in Eq.~(\ref{eq:Cspq}), which however has been ignored in previous studies~\citep{Porayko:2018sfa,NANOGrav:2023hvm,EuropeanPulsarTimingArray:2023qbc} and our analysis so far. As indicated in Eq.~(\ref{Eq:cov_term}), such a treatment trivializes the Earth and pulsar crossing terms also in the correlation functions. These terms  depend on the pulsar spatial position and exist for both auto-correlation and cross-correlation.

To look into the pulsar cross correlation, next let us extend the sensitivity analysis to the case where all terms in Eq.~\eqref{Eq:cov_term} are turned on. We consider J1231-1411 and J0614-3329, namely the two pulsars yielding the strongest upper limits individually, instead of using the full data of 28 pulsars as above. The pulsar spatial coordinates and their distance to the Earth are taken from the ATNF Pulsar Catalogue~\citep{Manchester:2004bp}. For the  marginalization in Eq.~\eqref{Eq:Psi_limit}, we take the  priors for pulsar distance introduced in Appendix~\ref{app:distance_prior}. We also assume no timing noise for simplifying the analysis, as its impacts on the exclusion limits tend to be tiny (see Fig.~\ref{Fig:exclusion1_DM} and Fig.~\ref{Fig:exclusion2_DM}). 
Three signal models are then comparatively analyzed: \textit{simplified correlation} where $(\mathcal{C}_a)_{pq}$ is simplified as Eq.~\eqref{Eq:Cspp}, \textit{auto-correlation only} where $(\mathcal{C}_a)_{p\neq q}$ terms are turned off in Eq.~\eqref{eq:Cspq} and \textit{full correlation} where all terms in Eq.~\eqref{eq:Cspq} are turned on.

    \begin{figure}[htb]
		\centering
        \includegraphics[scale=0.56]{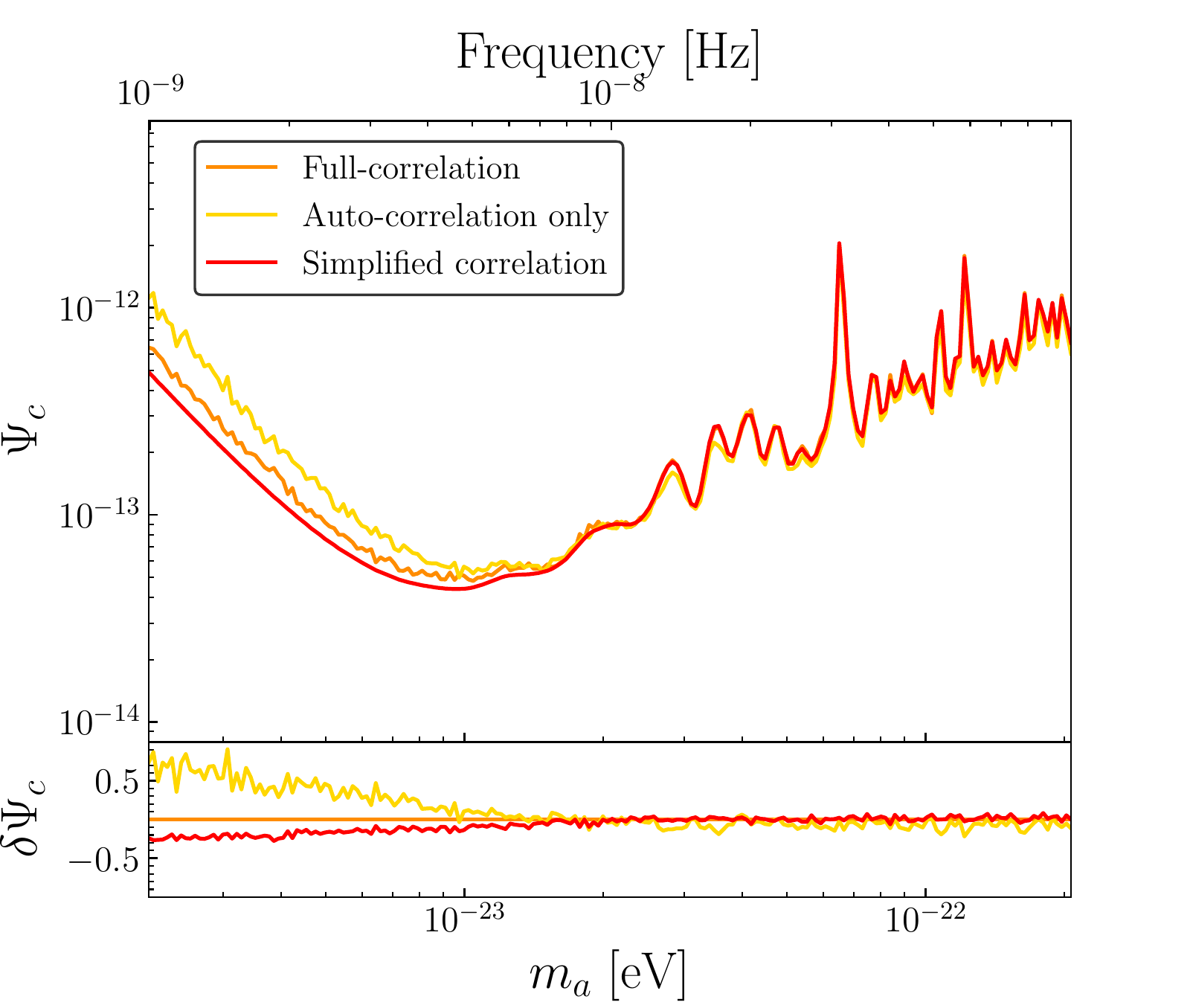}
        \caption{Upper limits of the Fermi-LAT $\gamma$-PTA (95\% C.L.) on $\Psi_c$, as a function of $m_a$. The limits are computed in three signal models: {\it simplified correlation}, {\it auto-correlation only} and {\it full-correlation}, with the Fermi-LAT data from the two pulsars yielding the strongest upper limits individually. The lower sub-panel shows the relative change of these limits with respect to the full-correlation analysis: $\delta\Psi_c = \Psi_c/\Psi_{c,{\rm full}}$ - 1.}
		\label{Fig:exclusion3_DM}
    \end{figure}

We demonstrate the upper limits on $\Psi_c$ in Fig.~\ref{Fig:exclusion3_DM}, as a function of $m_a$. The cross correlation plays a particularly important role for $m_a \lesssim 0.2 \, {\rm kpc}/v_0 \sim 10^{-23}\,$eV, {\it i.e.}, while the de Broglie wavelength of wave DM becomes bigger than the spatial separation between the two considered pulsars. Because of its contributions, the upper limits of \textit{full correlation} get visibly strengthened compared to those of \textit{auto-correlation}. We find this feature to be generic by performing the same analysis with the Fermi-LAT data of  various pulsar pairs. For larger $m_a$ values, all sinc function terms in Eq.~\eqref{Eq:cov_term} tend to be suppressed. Then, the contributions of auto- and cross-correlations to the covariance matrix and hence to the sensitivities become comparable. Notably, while the measurement of pulsar distance to the Earth is challenging, the information of cross- or full-correlations remains largely unaffected unless the distance uncertainties become larger than the de Broglie wavelength of wave DM\footnote{Note, as the GWs are relativistic, their de Broglie wavelength are about three orders of magnitude shorter than that of wave DM, for a given frequency. So, different from what occurs to the wave DM, the ``pulsar terms'' contributions to the PTA detection of stochastic GWs are subject to a strong suppression caused by the distance uncertainty of pulsars, for the frequency range of interest. For example, at nano-Hertz the “pulsar terms” can contribute coherently for the stochastic GWs only if the distance uncertainty is $\lesssim 1\,$pc, in comparison to $\lesssim 1\,$kpc for the wave DM.} (see Appendix~\ref{app:distance_prior} for details). 

\section{Discussion and Conclusion} 

Before concluding, let us take a review on the results demonstrated in last section. Let us start with a comparison of constraints between different radio PTA experiments. As shown in Fig.~\ref{Fig:exclusion1_DM} and Fig.~\ref{Fig:exclusion2_DM}, the EPTA constraints are stronger than the NANOGrav and PPTA ones in the low mass regime ($m_a\lesssim 1/T \sim 10^{-23}$~eV), and become comparable to them for the high mass regime ($m_a\gtrsim 1/T$). The gap in the low mass regime could be partly attributed to the longer data span of EPTA, which is 24.7 years, in comparison to NANOGrav's 15 years and PPTA's 12 years. As discussed in~\cite{EuropeanPulsarTimingArray:2023qbc}, such a difference may result in an enhancement of  constraints on $\Psi_c$ and $\rho_e$ by several times. At large masses, the longer data span of EPTA becomes less helpful in improving the constraints. The cadence instead becomes more relevant to the  detection and recognization of fast-oscillating signals from the dominant pulsar white noises~\citep{EuropeanPulsarTimingArray:2023qbc}.

Unlike these radio PTAs, the Fermi-LAT is not a dedicated PTA mission. But, the time span of 12.5 years for its pulsar observation~\citep{Fermi-LAT:2022wah} is only slightly shorter than that of NANOGrav. Particularly, the Fermi-LAT $\gamma$-PTA has collected the data uninterruptedly during its entire operation due to the constant experimental setup. Such a stability in data taking (and also an overall gain from the suppression of the intrinsic red noise of dispersion measure variance) makes the Fermi-LAT $\gamma$-PTA especially promising at masses below $10^{-23}$~eV~\citep{Fermi-LAT:2022wah}. The constraints comparable to the NANOGrav ones have thus been achieved in this mass regime,  
despite the telescope's much smaller collecting area. At high masses, however, the limitation stemming from the small number of photons becomes evident, resulting in a sensitivity gap between the Fermi-LAT $\gamma$-PTA and the radio PTAs which is enlarged gradually as $m_a$ increases.

Following the approach introduced in~\cite{Hazboun:2019vhv}, we can simply estimate the sensitivity prospects of detecting ultralight wave DM with the $\gamma$-PTA.  
In the low mass regime, the sensitivities are roughly scaled with $\rho_{e,95} \propto T^{-3}$. This implies that, by extending the observation time by another 12.5 years, the Fermi-LAT constraints could get close to the EPTA ones~\citep{EuropeanPulsarTimingArray:2023qbc} and begin ruling out a major contribution from wave DM. As for the high mass regime, the Fermi-LAT sensitivity will not improve significantly due to the dominance of white noise. To further improve the sensitivity, more dedicated $\gamma$-ray telescopes with significantly larger collecting areas and instant luminosities, either ground-based or spaceborne~\citep{Aharonian:2001, Fan:2022, Xie:2023}, are needed.

Finally, let us comment on the prospects of generalizing the \textit{full correlation} $\gamma$-PTA analysis from two to more pulsars. Unlike the \textit{simplified correlation} model, the calculation of likelihood function for the \textit{full correlation} model involves inverting the covariance matrix $\mathcal{C}_a$ with off-diagonal components. As the pulsar number increases, the complexity of this process grows quickly, as $\sim 8\mathcal{N}^3$. Furthermore, the dimension of parameter space becomes higher with more pulsars. More sampling points are thus required to fully explore the posterior distributions. Additionally, a more accurate analysis with red noise turned on would request more computing resources. It is therefore important to develop new likelihood calculation techniques for the future \textit{full correlation} data analysis with more pulsars. 

In summary, we have explored the potential of detecting the stochastic axion-like wave DM with the $\gamma$-PTA, using the recently released Fermi-LAT data. As it benefits from a suppression of the intrinsic red noise of dispersion measure variance, such a high-frequency PTA well-complements the traditional radio ones. Inspiringly, despite the much smaller collecting area of telescope, the upper limits generated for the energy density of axion-like wave DM are not far from those obtained in the more dedicated radio-PTA programs such as PPTA and NANOGrav, particularly for a mass range below the inverse of observation period. Such a strategic development could be extended to the PPA measurements in the near future. By then we would expect the hypothesis of wave DM to be examined gravitationally and non-gravitationally, in both radio and high-frequency bands. This will tremendously advance our exploration on the nature of DM.

{\bf [Note added]} While this Letter was being finalized, \cite{Xia:2023hov} appeared on arXiv. The application of $\gamma$-PTA for detecting ultralight wave DM was explored in both papers, with the Fermi-LAT data. However, this Letter distinguishes itself from \cite{Xia:2023hov} in terms of both signal nature (stochastic versus deterministic) and analysis method (PBP-based versus TOA-based). Particularly, a methodology to cross-correlate the data of arrayed pulsars, which is crucial for identifying the signal nature, has been developed and the relevant data analysis was fulfilled.  \\

\section*{Acknowledgements}
We would thank Jian Li for early-stage collaboration and Matthew Kerr for in-depth discussions and generous sharing of Fermi-LAT data and codes.
H. N. Luu, T. Liu and T. Broadhurst are supported by the Collaborative Research Fund under Grant No. C6017-20G which is issued by Research Grants Council of Hong Kong S.A.R.
J. Ren is supported by the Institute of High Energy Physics, Chinese Academy of Sciences, under Contract No. Y9291220K2.

\bibliography{reference}{}
\bibliographystyle{aasjournal}

\appendix

\section{Impacts of Pulsar Distance Uncertainty} \label{app:distance_prior}

As indicated in Eq.~(\ref{Eq:cov_term}), the information on pulsar spatial position is crucial for analyzing the cross-correlation  between the arrayed pulsars. Actually, the pulsar auto-correlation also relies on the pulsar distance, due to a crossing effect between the Earth and pulsar terms in Eq.~(\ref{Eq:phase_axion}). However, the measurement of pulsar distance is known to be challenging. Its uncertainty is typically $\sim \mathcal O(10\%)$ and even comparable to the measured value. Next let us look into the impacts of this uncertainty on the PTA sensitivity analysis.

For the Fermi-LAT pulsars, four types of distance measurements are relevant~\cite{NANOGrav:2023bts}, where the mean, uncertainty and prior of pulsar distance are determined by data~\citep{Manchester:2004bp}. The four types of distance measurements include:
\begin{itemize}

    \item A: the pulsar distance is measured with reference to other astronomical objects. In this case, the distance prior is given by a full Gaussian distribution with the mean $L_m$ and error $\delta L$, {\it i.e.},
    \begin{align}
        \ln \mathcal{L} \propto -\dfrac{(L - L_m)^2}{2\delta L^2}\,.
    \end{align}

    \item PX: the pulsar distance is measured through parallax observations, where the distance is given by the reciprocal of the parallax $\varpi$. In this case, the distance prior is derived from a full Gaussian distribution of the parallax, with the mean $\varpi_m$ and error $\delta \varpi$ 
    \begin{align}
        \ln \mathcal{L} \propto -\dfrac{(L^{-1} - \varpi_m)^2}{2\delta \varpi^2} -2\ln L \,.
    \end{align}
    
    \item DM: the pulsar distance, denoted as $L_{\rm DM}$ specifically, is measured through the dispersion measure, by using the YMW16 model of electron distribution~\citep{Yao:2017}. In this case, the distance prior is uniform for $0.8L_{\rm DM} \leq  L \leq 1.2L_{\rm DM}$ and follows a half Gaussian distribution otherwise, {\it i.e.}, 
\begin{align}
 \ln \mathcal{L} \propto 
 \left\{
\begin{array}{ll}
-\dfrac{(L - 0.8L_{\rm DM})^2}{2\delta L_{\rm DM}^2} \, ,  & \quad  L < 0.8 L_{\rm DM} \\
-\dfrac{(L - 1.2L_{\rm DM})^2}{2\delta L_{\rm DM}^2} \, , & \quad  L > 1.2 L_{\rm DM} \\
\end{array}\right.
\end{align}    
where the distance uncertainty is given by $\delta L_{\rm DM} = L_{\rm DM}/4$. 
    
    \item MM: only the lower bound $L_{\rm min}$ and upper bound  $L_{\rm max}$ of the distance are available. In this case, the distance prior is uniform for  $L_{\rm min} \leq  L \leq L_{\rm max}$ and vanishes otherwise.  
\end{itemize}
We present in Tab.~\ref{Tab:pulsar_distance} 
the information on the spatial coordinates and distance to the Earth for the nine pulsars in Fermi-LAT dataset~\citep{Fermi-LAT:2022wah} which yield the strongest constraints on $\Psi_c$ or $\rho_e$ individually. The data is from ATNF Pulsar Catalogue~\citep{atnf}, where some specific prioritization for the pulsar distance measurements has been taken. Concretely, the method ``PX'' is given the first precedence (if the mean parallax is three times larger than the quoted uncertainty) because it is the least dependent on a modeling assumption. The method ``A'' is prioritized next to ``PX'', which has a modest ``model dependence" because it is based on the absorption of neutral hydrogen from some astronomical objects. Different from the ``A'' and ``PX'' ones, the DM-derived distance 
is ``dependent'' on the model of electron distribution. So this method is relatively less accurate or solid than ``A'' and ``PX'' in general~\citep{atnf}. Finally, as the pulsar distance is unspecified in the method ``MM'', the relevant measurement is considered only if the measurements with the other methods are not available. Such a prioritization ensures that a more accurate or solid distance measurement for a given pulsar is more likely to be chosen. The information for J1231-1411 and J0614-3329 in this table has been applied for drawing Fig.~\ref{Fig:exclusion3_DM} and performing the relevant analysis. 

\begin{table}[htp]
\centering
\begin{tabular}{c|c|c|c|c}
	 Pulsar & Right ascension & Declination & Distance & Measurement \\
	\hline \hline
	J1231-1411 & 12:31:11.31 & -14:11:43.63 & $L_{\rm DM} = 0.42$ & DM \\
    J0614-3329 & 06:14:10.35 & -33:29:54.12 & $L = 0.54-0.63$ & MM \\
    J1959+2048 & 19:59:36.77 & 20:48:15.12 & $L = 1.4$ & A \\
    J0030+0451 & 00:30:27.43 & 04:51:39.71 & $\varpi = 3.09 \pm 0.06$ & PX \\
    J1630+3734 & 16:30:36.47 & 37:34:42.10 & $L_{\rm DM} = 1.187$ & DM \\
	J1614-2230 & 16:14:36.51 & -22:30:31.30 & $\varpi = 1.54 \pm 0.10$ & PX \\
    J1939+2134 & 19:39:38.56 & 21:34:59.12 & $\varpi = 0.21 \pm 0.05$ & PX \\
	J1902-5105 & 19:02:02.85 & -51:05:56.97 & $L_{\rm DM} = 1.645$ & DM \\
	J2302+4442 &  23:02:46.98 & 44:42:22.08 & $L_{\rm DM} = 0.863$ & DM \\
\end{tabular}
\caption{Spatial coordinates and distance to the Earth for the nine pulsars in Fermi-LAT dataset~\citep{Fermi-LAT:2022wah} which yield the strongest constraints on $\Psi_c$ or $\rho_e$ individually. The distance and parallax are given in unit of kpc and kpc$^{-1}$, respectively. The data is taken from ATNF Pulsar Catalogue~\citep{atnf}. In the case where the distance for a given pulsar has been measured in multiple ways, we consider its measurement following a priority order: PX, A, DM, and MM. The distance uncertainty of J1959+2048 is undetermined~\citep{Brownsberger:2014}.
}
\label{Tab:pulsar_distance}
\end{table}

Formally, the uncertainty of pulsar distance influences the correlation functions via triangular modulation and sinc function, which are characterized by the Compton wavelength and de Broglie wavelength (or coherence length) of the axion-like wave DM, respectively (see Eq.~(\ref{Eq:cov_term})). These two length scales are distinguished by a factor of $v_0^{-1} \sim 10^3$. 
When the pulsar distance is much larger than the de Broglie wavelength, {\it i.e.}, $y_{ij}\gg 1$, the relevant terms are suppressed by the sinc function. The uncertainty of pulsar distance becomes irrelevant for the sensitivity analysis then. In this discussion, let us focus on the case with $y_{ij}\lesssim 1$ where the sinc function is approximately equal to one. While the uncertainty of pulsar distance becomes much bigger than the Compton wavelength, the triangular modulation starts to oscillate rapidly as the distance varies within the prior range. Such a rapid oscillation, although yielding an average $\langle \sin[2m_aL_{p, pq}] \rangle = \langle \sin[2m_aL_{p, pq}]\rangle \approx 0 $, does not fully suppress the contributions of triangular modulation to the marginal likelihood $\mathcal{L}_m$ in Eq.~(\ref{Eq:Psi_limit}). As the likelihood $\mathcal{L}$ is an exponential functional of the correlation functions in Eq.~(\ref{Eq:cov_term}), marginalizing the triangular modulation over the pulsar distance yields a deviation of $\mathcal O(0.1)$ from one in $\mathcal{L}_m$ effectively. If the distance uncertainty becomes even much bigger than the de Broglie wavelength, the marginalization of sinc function over the pulsar distance gets highly suppressed. The marginal likelihood $\mathcal{L}_m$ for the full correlation is then reduced to the one for the simplified correlation, where all sinc function terms in Eq.~\eqref{Eq:cov_term} are turned off.

\begin{figure}[htb]
	\centering
 \includegraphics[scale=0.55]{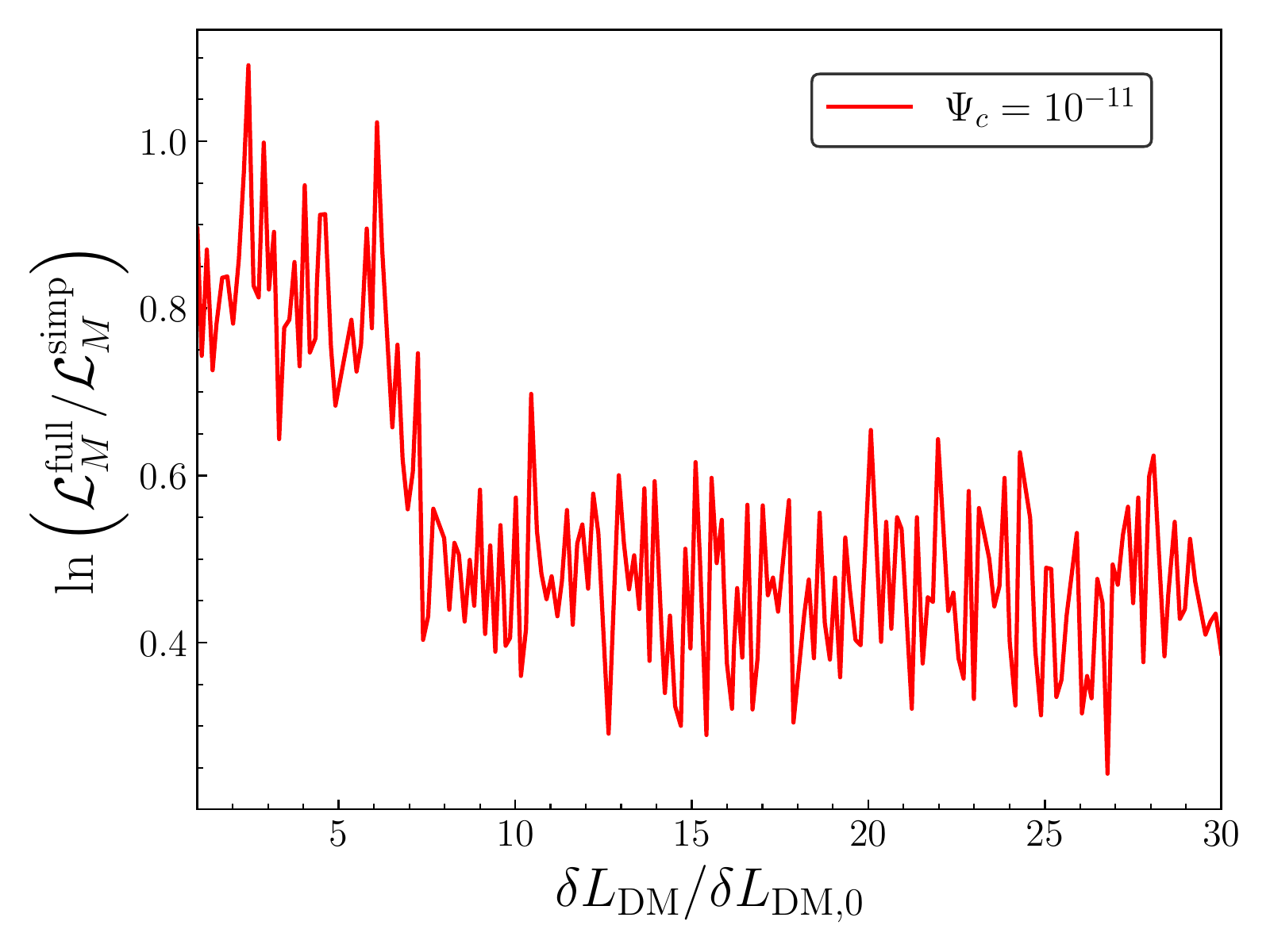}
        \includegraphics[scale=0.55]{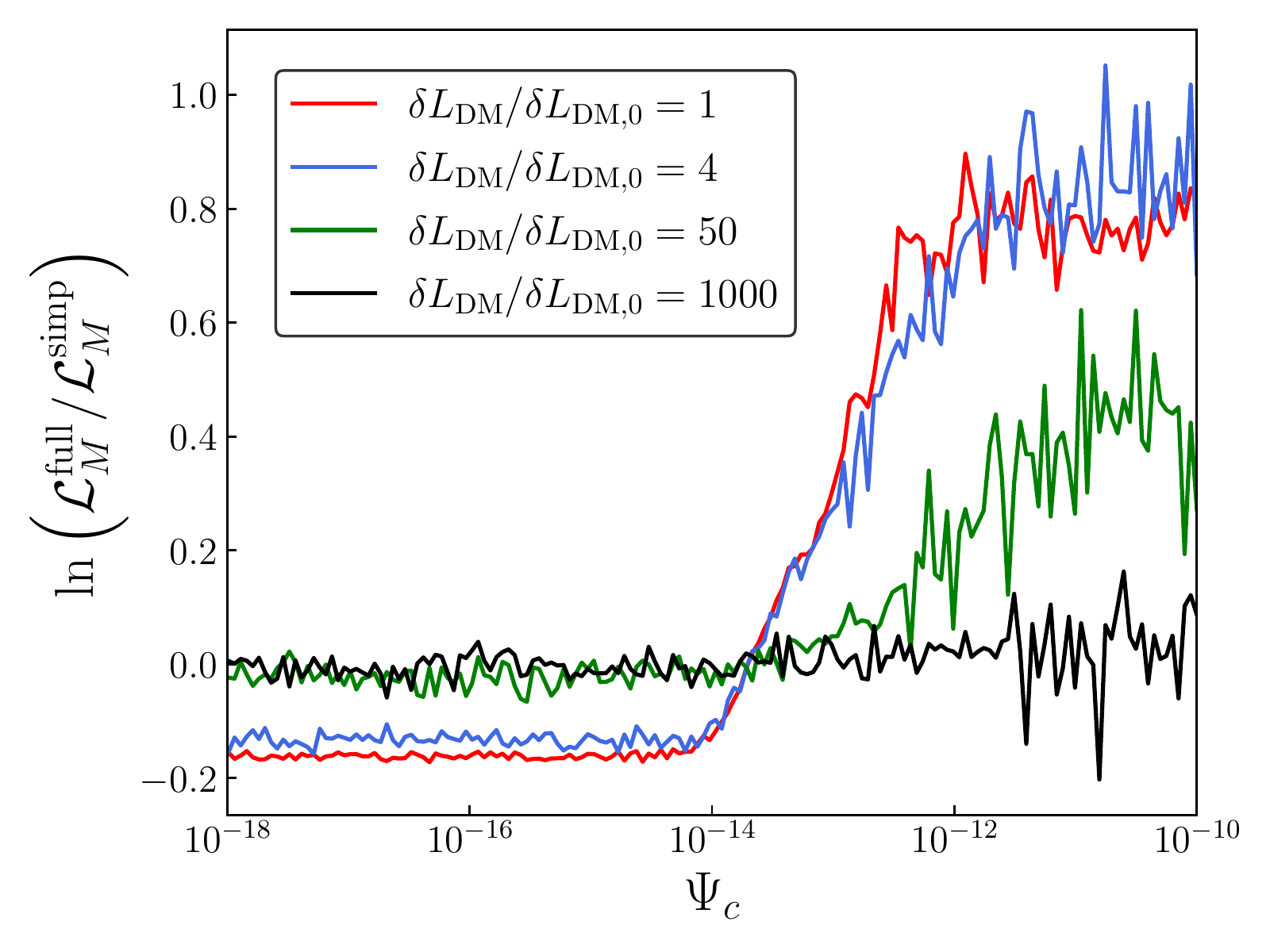} 
    \caption{Logarithm of the marginal likelihood ratio between the full and simplified correlation models ($\ln (\mathcal{L}_m^{\rm full}/ \mathcal{L}_m^{\rm simp})$), as a function of $\delta L_{\rm DM}/\delta L_{0,\rm DM}$ for $\Psi_c = 10^{-11}$ (left) and of $\Psi_c$ for varied $\delta L_{\rm DM}/\delta L_{0,\rm DM}$ values (right). Here $m_a=5\times 10^{-24}\,$eV has been assumed. The two pulsars selected, namely J1231-1411 and J1630+3734, have a spatial separation of  $1.16\,$kpc.}

	\label{Fig:exclusion_sigma}
\end{figure}

To visualize this point, we demonstrate $\ln (\mathcal{L}_m^{\rm full}/ \mathcal{L}_m^{\rm simp})$ as a function of the dimensionless measure $\Psi_c$ and the distance uncertainty in Fig.~\ref{Fig:exclusion_sigma}, for $m_a = 5 \times 10^{-24}\,$eV and with two pulsars. For the convenience of discussion, we have considered J1231-1411 and J1630+3734, while the conclusion is generic and general. These two pulsars have a spatial separation 1.16\,kpc, which is slightly shorter than the de Broglie wavelength $l_c \approx 1.3\,$kpc. This is also true for their distances to the Earth. As both distances are measured with the DM method, we can vary their uncertainties universally by changing the value of $\delta L_{\rm DM}/\delta L_{0,\rm DM}$. Here $\delta L_{0,\rm DM}=L_{\rm DM}/4$ denotes the original uncertainty. Clearly, these two signal models are distinguishable for the original distance uncertainties, which are significantly larger than the Compton wavelength for both pulsars. Such a situation keeps nearly unchanged as the distance uncertainty increases until it becomes comparable to the de Broglie wavelength. As shown in the left panel, the $\ln (\mathcal{L}_m^{\rm full}/ \mathcal{L}_m^{\rm simp})$ value jumps down at $\delta L_{\rm DM}/\delta L_{0,\rm DM} \sim 6-7$, which exactly encodes the effect discussed above. Accordingly, the curve in the right panel starts to shift downwards significantly. It becomes an approximately constant curve with $\ln (\mathcal{L}_m^{\rm full}/ \mathcal{L}_m^{\rm simp}) \approx 0$ when $\delta L_{\rm DM}$ becomes extremely large. This implies that the information carried by the distance-related terms in the correlation functions has been smeared by the distance uncertainty. These discussions highlight the importance of including the distance-related terms in Eq.~\eqref{Eq:cov_term}, particularly the ones characterizing the cross correlation, which have been universally neglected in the previous studies on deterministic signals~\citep{Porayko:2018sfa,NANOGrav:2023hvm,EuropeanPulsarTimingArray:2023qbc}.

\end{document}